\documentclass[sigconf]{acmart}

\AtBeginDocument{%
  }

\acmConference[ICSE 2024]{46th International Conference on Software Engineering}{April 2024}{Lisbon, Portugal}

\usepackage{amsmath}
\usepackage[ruled]{algorithm2e}
\usepackage{booktabs}
\usepackage{subfig}
\usepackage{bm}

\begin{document}

\title{Contrastive Prompt Learning-based Code Search based on Interaction Matrix}

\author{Yubo Zhang}
\email{yubozhang@buaa.edu.cn}
\affiliation{
  \institution{School of Computer Science and Engineering, Beihang University}
  \city{Beijing}
  \country{China}
}

\author{Yanfang Liu}
\email{hannahlyf@buaa.edu.cn}
\affiliation{
  \institution{School of Computer Science and Engineering, Beihang University}
  \city{Beijing}
  \country{China}
}

\author{Xinxin Fan}
\email{fanxinxin@ict.ac.cn}
\affiliation{
  \institution{Institute of Computing Technology, Chinese Academy of Sciences}
  \city{Beijing}
  \country{China}
}

\author{Yunfeng Lu}
\authornote{Corresponding author.}
\email{lyf@buaa.edu.cn}
\affiliation{
  \institution{School of Reliability and Systems Engineering, Beihang University}
  \city{Beijing}
  \country{China}
}

\begin{abstract}
Code search aims to retrieve the code snippet that highly matches the given query described in natural language. Recently, many code pre-training approaches have demonstrated impressive  performance on code search. However, existing code search methods still suffer from two performance constraints: inadequate semantic representation and the semantic gap between natural language (NL) and programming language (PL). In this paper, we propose CPLCS, a contrastive prompt learning-based code search method based on the cross-modal interaction mechanism. CPLCS comprises:(1) PL-NL contrastive learning, which learns the semantic matching relationship between PL and NL representations; (2) a prompt learning design for a dual-encoder structure that can alleviate the problem of inadequate semantic representation; (3) a cross-modal interaction mechanism to enhance the fine-grained mapping between NL and PL. We conduct extensive experiments to evaluate the effectiveness of our approach on a real-world dataset across six programming languages. The experiment results demonstrate the efficacy of our approach in improving semantic representation quality and mapping ability between PL and NL. 
\end{abstract}



\maketitle

\section{Introduction}
Code search involves searching and retrieving relevant code snippets from a code repository based on the textual intent expressed by the developer in a search query. Today, with the rise of large code projects and advanced search capabilities, code search has become a crucial software development activity. It also supports many other important software engineering tasks such as locating and addressing software bugs and defects \cite{1}, repairing programs \cite{2}, synthesizing code \cite{3}, and detecting vulnerabilities \cite{4}.

Initially, code search models \cite{22,23} relied on traditional information retrieval (IR) techniques, such as keyword matching, or application program interface (API) matching. Later on, researchers have demonstrated that code pre-training techniques, such as CodeBERT \cite{6} and GraphCodeBERT \cite{7}, could dramatically enhance code search performance by leveraging large-scale code corpus \cite{8} through self-supervised pre-training. Furthermore, several pre-training methods based on contrastive learning such as Corder \cite{16}, Code-MVP \cite{11}, CodeRetriver \cite{14}, SynCoBERT \cite{12}, and CoCoSoDa \cite{10} have been proposed to obtain better code representations. The underlying principle of contrastive learning is to bring similar samples together while pushing dissimilar samples apart. This line of work maps both natural language (NL) and programming language (PL) into the same semantic space, and then retrieves the corresponding code based on similarity between the input NL query and the candidate PL snippets. 

However, these code pre-training approaches usually employ masked language modeling (MLM) as one of the pre-training tasks, which often produces inadequate semantic representations of PL and NL \cite{14}. Some researchers attribute the issue to anisotropy \cite{15,17}, while others \cite{18,20} argue that it may be due to static word embedding bias or ineffective Transformer layers. Although the reason is still under debate, inadequate representations will ultimately lead to poor performance in the tasks based on representational similarity, such as code search.

Additionally, due to the semantic gap across modalities, it is more challenging to establish semantic mappings between NL and PL compared to mappings between texts. Current mainstream approaches all employ the dual-encoder structure, aiming to utilize the dual-encoders that share weights in learning both intra- and inter-modal relationships simultaneously. However, accomplishing this learning in practice can often be challenging as the dual-encoder structure does not explicitly model cross-modal information.

Recent studies \cite{5,21} have demonstrated the impressive performance of prompt learning in code intelligence applications, particularly when working in low-resource scenarios. Therefore, we propose CPLCS (\textbf{C}ontrastive \textbf{P}rompt \textbf{L}earning Code Search with \textbf{C}ross-modal Interaction Mechani\textbf{s}m) and introduce the reparameterization mechanism before the input embedding layer of each encoder. Considering the heterogeneity of NL and PL, we refrain from weight sharing between reparameterization encoders. Taking inspiration from ColBERT \cite{37}, we design a lightweight cross-modal interaction mechanism for code search. This mechanism facilitates fine-grained mapping between NL and PL while still preserving the ability to pre-compute NL and PL representations offline. Concretely, our work makes the three contributions as follows:

\begin{itemize}
\item To the best of our knowledge, we are the first to propose prompt learning for the dual-encoder structure to optimize multimodal code contrastive learning, which addresses the problem of inadequate representation from a fresh perspective, and can achieve faster training and less resource consumption.
\item We propose a lightweight cross-modal interaction mechanism to facilitate fine-grained mapping between NL and PL, so as to bridge the semantic gap.
\item We conduct extensive experiments to evaluate the proposed approach. Results demonstrate that our approach significantly outperforms the baseline models on most PLs. 
\end{itemize}

The rest of this paper is organized as follows. Section 2 reviews the related work, and Section 3 details our approach. Section 4 presents the experiment setup and analyzes the experimental results. Finally, we conclude this paper in Section 5.

\section{Related Work}
\subsection{Code Search}
Initially, code search models relied on traditional information retrieval (IR) techniques, such as keyword matching \cite{22}, or a combination of text similarity and application program interface (API) matching \cite{23}. However, their retrieval accuracy is unstable, and the application scenarios are limited.

Nowadays, the code search is based on the model of dual-encoder architecture, where NL queries and PL code segments are separately fed into two encoders with shared parameters. The encoders' outputs pass through their respective feature extraction layers to derive query embedding and code embedding. The similarity between the query and the code is subsequently determined by a matching function. For example, Gu et al. developed the first deep learning model DeepCS \cite{24} for code search, which utilizes LSTM to map PL and NL into a unified semantic space, and incorporates the additional information such as API sequences and function names to encode code semantics alongside code tokens. Yao et al. proposed the MRNCS method \cite{25} to convert the abstract syntax tree of the code into a simplified semantic tree (SST). The SST is then utilized as additional information alongside the source code for joint encoding to obtain the code representation vector.

Recently, a number of pre-training code language models have emerged. CodeBERT utilizes MLM and replace word detection (RTD) tasks during pre-training. It supports six different languages and has demonstrated effective scalability. GraphCodeBERT is built on the basis of CodeBERT by further utilizing the code data flow to enhance code search efficiency. Code-MVP learns representations through contrastive pre-training, using multiple views of the code and its semantic preserving forms. CoCoSoDa employs soft data augmentation techniques to construct positive samples for contrastive learning, and utilizes momentum encoders to stabilize the learning process of the model.

Unlike aforementioned studies, our model utilizes multimodal contrastive learning by introducing prompt tuning and lightweight cross-modal interaction mechanism to learn better representations, which leads to superior performance in code search tasks.

\subsection{Semantic Representation in NLP}

Semantic representation is essential for search tasks in NLP, but the currently popular models are deficient in semantic representations. Gao et al. \cite{31} and Wang et al. \cite{32} discovered that word embeddings learned by Transformer-based neural network models, such as BERT and GPT, exhibit a tendency to degenerate and cluster within a narrow cone. This phenomenon significantly restricts the representational expressiveness of word embeddings, and thus adversely impacts the models’ performance.

Li et al. \cite{33} attributed the previously mentioned phenomenon to anisotropy, and proposed BERT-flow to correct the distribution of pre-trained word vectors by using the flow model. Su et al. \cite{34} further proposed an efficient method BERT-whitening, which employs a simple linear transformation instead of the flow model. Other studies have attempted to utilize contrastive learning to mitigate the inadequate semantic representation such as SimCSE \cite{35} and ConSERT \cite{36}. Wang et al. \cite{19} theoretically proved that the InfoNCE loss function used by contrastive learning can help align and distribute word vectors more consistently, resulting in more adequate representation. Jiang et al. \cite{18}, on the other hand, argue that the subpar performance of BERT in similarity tasks is mainly due to the bias of static tokens and invalid BERT layers. Baggetto et al. \cite{20} further conducted additional experiments to support Jiang’s view.

Although the reason of inadequate representations is still under debate, the manifestations are all in the form of either a too narrow distribution of the token embedding learned by the model in the semantic space, or relatively large distances between similar samples. Consequently, this leads to poor performance in the tasks based on representational similarity, such as code search.

\subsection{Prompt-based Learning}
Prompt-based learning approaches have gained popularity as a promising method in low-resource scenarios\cite{26}. For instance, Schick et al. \cite{27} proposed PET, which transforms the classification task into a MLM task and leverages prompt to elicit knowledge from a pre-trained language model (PLM). However, manually crafting prompts and selecting the most suitable words can be challenging.

Since the neural model is inherently continuous, it can never reach the optimum with discrete prompt tokens from an optimization point of view. To address this issue, Liu et al. proposed P-tuning \cite{28} that enables automatic search of knowledge templates in continuous space. P-tuning freezes the PLM, and only trains the template itself to reduce computational costs. However, this approach often results in severe degradation of models with fewer than 10B parameters. Prefix-Tuning proposed by Li et al. \cite{30} optimizes a continuous task-specific vector attached to every layer of the Transformer in PLM, demonstrating impressive performance. Nonetheless, this method is limited to NL generation tasks. Inspired by Prefix-tuning, Liu et al. subsequently proposed P-tuning v2\cite{29} and extended Prefix-Tuning to the fields of natural language understanding (NLU). The performance of P-Tuning v2 is comparable with fine-tuning on PLMs ranging in size from 300 M to 10 B parameters. In our work, we further modify P-tuning v2 to accommodate the dual-encoder structure model in code search task.

\begin{figure*}[htp]
  \centering
  \includegraphics[scale=0.22]{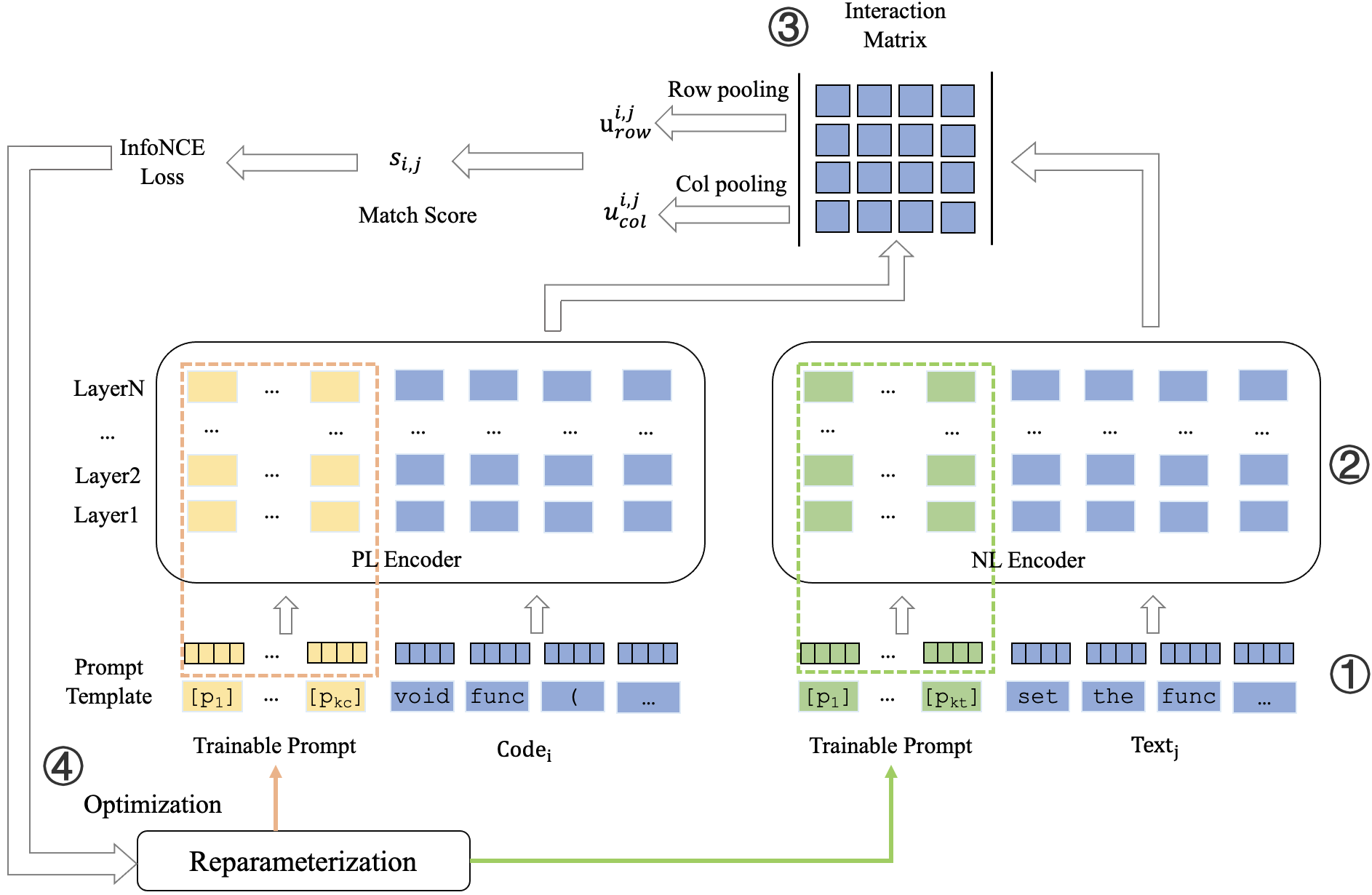}
  \caption{Model training workflow.}
  \label{fig:1}
\end{figure*}

\section{Methodology}
\subsection{Overview}
In this section, we first present the basic framework of our approach CPLCS. Figure \ref{fig:1} illustrates the training flow of our proposed approach. We begin by constructing prompts and task-specific vectors based on the given code snippet and its corresponding text using the prompt template.(corresponds to \textcircled{1} in the Figure) Then, we input the PL and NL with their prompts into the respective encoder and learn the token-level embedding through contrastive prompt tuning.(\textcircled{2}) Next, we construct an interaction matrix based on the similarities between each pair of tokens and perform a pooling operation on the matrix to obtain the match score.(\textcircled{3}) Finally, we calculate the loss function based on these match scores and optimize the prompt template with reparameterization.(\textcircled{4})

\subsection{Prompt Tuning for Dual-encoder Structure}
Prompt tuning is a technique that involves fixing all parameters of a pre-trained model and tuning only the continuous prompt template. Studies \cite{5,21} have shown that prompt tuning can yield comparable results to fine-tuning results, while requiring fewer resources for code summarization and clone detection.

As illustrated in Figure \ref{fig:1}, we propose a novel prompt tuning method for dual-encoder structures in code search. First, we eliminate weight sharing between the PL encoder and NL encoder. Next, we utilize a prompt template to generate prompt and task-specific vectors for each encoder. The process of generating continuous prompts and task-specific vectors can be described as follows:
\begin{equation}\label{equation_1}
P_{PL},V_{PL}\leftarrow G(PL;\phi_X )
\end{equation}
\begin{equation}\label{equation_2}
P_{NL},V_{NL}\leftarrow G(NL;\phi_Y )
\end{equation}
where $\phi_X$ and $\phi_Y$ are the parameters of the trainable prompt templates on the PL and NL sides, respectively, and serve as the final optimization objectives; $P_{PL}=\{p_{1},...,p_{kc}\}$ and $P_{NL}=\{p_{1},...,p_{kt}\}$ represent the continuous prompts of PL and NL, respectively; $kc$ and $kt$ correspond to the prompt lengths of PL and NL, respectively; $V_{PL}$ and $V_{NL}$ are the task-specific vectors of PL and NL with the shapes $[kc,ln,emb]$ and $[kt,ln,emb]$, where $ln$ is the number of Transformer layers in the encoder, and $emb$ is the embedding size of the model. Moreover, in order to enhance the training stability, we also leverage MLP as a reparameterization encoder to transform continuous prompts. The process of reparameterization can be described as follows:
\begin{equation}\label{equation_3}
P_{PL},V_{PL}\leftarrow H(P_{PL},V_{PL};\psi_{X} )
\end{equation}
\begin{equation}\label{equation_4}
P_{NL},V_{NL}\leftarrow H(P_{NL},V_{NL};\psi_{Y} )
\end{equation}
where $\psi_X$ and $\psi_Y$ are the parameters of the trainable feed forward network on the PL and NL sides, and the input and output dimensions of the reparameterization encoder are the same.

Finally, we reconstruct PL and NL by adding the prompts as follows:
\begin{equation}\label{equation_5}
X\leftarrow \{P_{PL},[CLS],PL,[SEP]\}
\end{equation}
\begin{equation}\label{equation_6}
Y\leftarrow \{P_{NL},[CLS],NL,[SEP]\}
\end{equation}
where $X$ and $Y$ here are the final inputs to the encoder.

\subsection{PL-NL Contrastive Learning with Cross-modal Interaction Mechanism}
Wang et al. \cite{19} proved that optimizing the InfoNCE loss function is equivalent to optimizing alignment and uniformity in hyperspherical semantic space, and can contribute to a more adequate semantic representation. The dual-encoder architecture also facilitates the transformation between different modal data and is thus widely used in multimodal scenarios such as code-text and image-text.

In general, we employ multimodal contrastive learning for training, and use instance discrimination as the pretext task. However, unlike previous works \cite{10,11,12}, which directly utilize cosine similarity to measure the distance between samples, we propose the cross-modal interaction mechanism to compute the similarity between samples, and further enhance the fine-grained mapping between PL and NL.

Specifically, we first input the prompt alongside <PL,NL> pair into the dual-encoder to obtain the token-level representations of the code $\bm{r_X}$ and text $\bm{r_Y}$:
\begin{equation}\label{equation_7}
\bm{r_X}\leftarrow F(X,V_{PL};\theta _X )
\end{equation}
\begin{equation}\label{equation_8}
\bm{r_Y}\leftarrow F(Y,V_{NL};\theta _Y )
\end{equation}
where $\bm{r_X}=\{\bm{x_1},…,\bm{x_{mc}}\}$, $\bm{r_Y}=\{\bm{y_1},..,\bm{y_{mt}}\}$, and $mc$ and $mt$ represent the maximum sequence length of PL and NL, respectively; $\theta_X$ and $\theta_Y$ are the parameters of PL and NL encoders, which are frozen in the training phase; $V_{PL}$ and $V_{NL}$ are task-specific vectors, which are the left values in Equation \ref{equation_3} and \ref{equation_4}.

Assuming a batch size of $N$ in the characterization of the set $\{(\bm{r_X^i},\bm{r_Y^j})\}_{i,j=1}^{N}$ to compute the similarity between the $i$th PL sample and the $j$th NL sample, we first need to construct the interaction matrix according to the following equation:
\begin{equation}\label{equation_9}
M^{i,j}= \bm{r_X^i} \cdot \bm{{r_Y^j}}^T
\end{equation}

Next, we perform maximum pooling operation on both the rows and columns of the matrix to obtain the two factors as follows:
\begin{equation}\label{equation_10}
u_{row}^{i,j}=\frac{1}{mc}\sum_{k=1}^{mc}max(M_{k,:}^{i,j})
\end{equation}
\begin{equation}\label{equation_11}
u_{col}^{i,j}=\frac{1}{mt}\sum_{k=1}^{mt}max(M_{:,k}^{i,j})
\end{equation}

Finally, according to the above two factors, we compute the similarity between the $i$th code sample and the $j$th text sample $s_{i,j}$ as follows:
\begin{equation}\label{equation_12}
s_{i,j}=\lambda u_{col}^{i,j}+(1-\lambda) u_{row}^{i,j}
\end{equation}
where $\lambda$ is a hyperparameter set manually. Let $\mathcal{L}_{X2Y}$ and $\mathcal{L}_{Y2X}$ represent respectively the PL-to-NL loss and NL-to-PL loss. Given a batch of reconstructed PLs and NLs, the loss $\mathcal{L}_{X2Y}$ is defined to match each PL to its corresponding NL as follows:
\begin{equation}\label{equation_13}
\mathcal{L}_{Y2X}=-\sum_{i=1}^{N}log\frac{exp(\frac{s_{i,i}}{\tau})}{\sum_{j=1}^{N} {exp(\frac{s_{i,j}}{\tau})}}
\end{equation}
where $\tau$ is the temperature parameter. Similarly, we have $\mathcal{L}_{Y2X}$ to match each NL to its corresponding PL as follows:
\begin{equation}\label{equation_14}
\mathcal{L}_{Y2X}=-\sum_{i=1}^{N}log\frac{exp(\frac{s_{i,i}}{\tau})}{\sum_{j=1}^{N} {exp(\frac{s_{j,i}}{\tau})}}
\end{equation}
To this end, the overall multimodal contrastive learning loss function for a batch is:
\begin{equation}\label{equation_15}
L_{(X,Y)\in B}=\frac{1}{2N}(\mathcal{L}_{X2Y}+\mathcal{L}_{Y2X} )
\end{equation}

\begin{algorithm}[]
    \caption{Contrastive Prompt Tuning with Cross-modal Interaction Mechanism}
    \LinesNumbered 
        \KwIn{<PL,NL> pairs datasets $D$ comprised of batches}
        \KwOut{output result}
        Initialization of $F_{\theta_X}$ and $F_{\theta_Y}$ with CoCoSoDa pre-trained parameters and freeze $\theta_X$,$\theta_Y$\;
        Initialization of prompt template for PL and NL: $G_{\phi_X}$, $G_{\phi_Y}$\;
        Initialization of reparameterization encoder for PL and NL: $H_{\psi_X}$, $H_{\psi_Y}$\;
        \For{Each batch $B \in D$}{
        \For{Each $PL, NL \in B$}{
            Generate continuous prompts and task-specific vectors based on template:
            
            $P_{PL},V_{PL}\leftarrow G(PL;\phi_X )$ 
            
            $P_{NL},V_{NL}\leftarrow G(NL;\phi_Y )$ 
            
            Conduct the reparameterization process:
            
            $P_{PL},V_{PL}\leftarrow H(P_{PL},V_{PL};\psi_{X} )$ 
            
            $P_{NL},V_{NL}\leftarrow H(P_{NL},V_{NL};\psi_{Y} )$ 

            Reconstruct PL and NL by adding prompts:

            $X\leftarrow \{P_{PL},[CLS],PL,[SEP]\}$
            
            $Y\leftarrow \{P_{NL},[CLS],NL,[SEP]\}$
            
            $\bm{r_X}\leftarrow F(X,V_{PL};\theta _X )$
            
            $\bm{r_Y}\leftarrow F(Y,V_{NL};\theta _Y )$
        }
        Construct interaction matrix $M$ based on $\{(\bm{r_X^i},\bm{r_Y^j})\}_{i,j=1}^{N}$

        Compute match scores based on $M$ of PLs and NLs in the batch $B$

        Compute symmetric loss with $\mathcal{L}_{X2Y}$, $\mathcal{L}_{Y2X}$: $L_{(X,Y)\in B}=\frac{1}{2N}(\mathcal{L}_{X2Y}+\mathcal{L}_{Y2X} )$

        Joint optimization of the symmetric objective: $argmin_{\{\phi_X,\phi_Y,\psi_X,\psi_Y\}} \mathcal{L}_{(X,Y)\in B}$
        }
\end{algorithm}

The Contrastive Prompt Tuning with Cross-modal Interaction Mechanism training process can be summarized as Algorithm 1. With the prompt tuning process in Lines 12-16 and 21, we reconstruct PL and NL with the generated prompts and update only the parameters of continuous prompt and reparameterization encoders. As demonstrated in Lines 18-20 of the training process, we employ the cross-modal interaction mechanism and symmetric objective function for dual-encoders to learn better embeddings.

\section{Experiment}
\subsection{Experiment Configuration}

\textbf{Dataset.} We conduct experiments on a large-scale benchmark dataset CodeSearchNet \cite{8}, which encompasses six programming languages: Ruby, JavaScript, Go, Python, Java and PHP. To enhance representativeness of real-world search scenarios, Guo et al. \cite{7} proposed a refined candidate set based on the original CodeSearchNet. Consistent with prior studies \cite{7,10,24,38}, we performed experiments on the dataset revised by Guo et al. \cite{7}, and the statistics are presented in Table \ref{tab:1}.

\begin{table}[]
\caption{Dataset statistics}
\begin{tabular}{lllll}
\hline
Language   & Training & Validation & Test   & Candidate \\ \hline
Ruby       & 24,927   & 1,400      & 1,261  & 4,360     \\
JavaScript & 58,025   & 3,885      & 3,291  & 13,981    \\
Java       & 164,923  & 5,183      & 10,955 & 40,347    \\
Go         & 167,288  & 7,325      & 8,122  & 28,120    \\
PHP        & 241,241  & 12,982     & 14,014 & 52,660    \\
Python     & 251,820  & 13,914     & 14,918 & 43,827    \\ \hline
\end{tabular}
\label{tab:1}
\end{table}

\textbf{Baseline.} We compare our model with seven pre-training-based approaches, including two early pre-trained models, one prompt learning-related model and four contrastive learning-related models.
\begin{itemize}
\setlength{\leftmargin}{0pt}
\item CodeBERT is built based on Transformer and is pre-trained with both NL and PL on a large code corpus with MLM and RTD tasks.
\item GraphCodeBERT improves CodeBERT by additionally integrating code structure information during pre-training. Specifically, it is pre-trained through data flow edge prediction, and node alignment tasks.
\item Zecoler tackles downstream tasks by using trainable prompts to reformat them into the same forms as those of the pre-trained tasks. The pre-trained model is then optimized by adjusting only the raw input. This method has proven to be effective in code understanding scenarios, especially in low-resource scenarios.
\item SyncoBERT aims to leverage the complementary information from semantically equivalent modals, such as source code, comments and AST, to learn the lexical and syntactical knowledge of source code. This is achieved through contrastive pre-training, whereby identifier and AST edge predictions are learned.
\item UniXcoder fully leverages the code structure information provided by AST, using flattened AST and source code data as inputs. It is pre-trained through multiple tasks, including MLM, unidirectional language modeling (ULM), denoising autoencoder, and two contrastive learning tasks.
\item CodeRetriever uses unimodal and bimodal contrastive learning approaches for learning function and code representation, as well as mixed language representation. To achieve this goal, a large training corpus of both PL-PL and PL-NL is constructed.
\item CoCoSoda proposes four dynamic soft data enhancement methods and momentum encoders to improve the effectiveness of multimodal contrastive learning. It uses both unimodal and bimodal contrastive learning during pre-training, similar to CodeRetriver.
\end{itemize}

\textbf{Evaluation Metrics.} We use three widely used metrics to evaluate the code search methods: FRank, MRR and Recall@k. \textbf{FRank}, or best hit rank, is the rank of the first hit result in the result list. A smaller FRank implies a lower inspection effort to find the desired result. Mean Reciprocal Rank (\textbf{MRR}) is the average of the reciprocal ranks of FRank of the query set, and is another popular evaluation metric for the code search task. The metric MRR is calculated using the equation as follows:
\begin{equation}\label{equation_16}
MRR=\frac{1}{|Q|}\sum_{i=1}^{|Q|}\frac{1}{FRank_{i}}
\end{equation}
\textbf{Recall@k} calculates the proportion of queries that the relevant code could be found in the top-k ranking list. It is calculated as follows:
\begin{equation}\label{equation_17}
Recall@k=\frac{1}{|Q|}\sum_{i=1}^{|Q|} \delta(FRank_i\leq k)
\end{equation}
where $\delta$ is an indicator function that returns 1 if $FRank_i\leq k$;  otherwise returns 0. In our settings, we take $k$ as 1, 5, and 10. We use R@k to refer to Recall@k in the following.

\textbf{Implementation Details.} We implement the dual-encoder using Transformer with 12 layers, 768 dimensional hidden states, and 12 attention heads. The dual-encoder is initialized with CoCoSoDa pre-trained parameters. We set the maximum length of PL and NL to 256 and 128, respectively. Our experimental implementation is based on the Huggingface Transformers\footnote{\url{https://www.huggingface.co/DeepSoftwareAnalytics/CoCoSoDa}} and PEFT\footnote{\url{https://github.com/huggingface/peft}}. The batch size, learning rate, temperature hyperparameter and the number of epochs are set to 160, 2e-5, 0.05 and 10, respectively. We freeze the weight of the original PL and NL encoders, cancel the weight sharing between reparameterization encoders and prompt templates for PL and NL. The model is optimized using the AdamW algorithm on a machine with a Nvdia A40 48GB GPU. The average results are reported under 3 different random seeds.

\subsection{Performance Evaluation}
In this section, we answer the following several research questions (RQs) to validate the effectiveness of our proposed approach.

\begin{table*}[htbp]
\caption{Performance of different approaches on CodeSearchNet}
\label{tab:2}
\begin{tabular}{llllllll}
\hline
Model         & Ruby  & JavaScript & Go    & Python & Java           & PHP   & Overall \\ \hline
CodeBERT      & 0.678 & 0.622      & 0.879 & 0.670  & 0.677          & 0.625 & 0.692   \\
GraphCodeBERT & 0.703 & 0.644      & 0.895 & 0.693  & 0.686          & 0.649 & 0.712   \\ \hline
Zecoler       & 0.701 & 0.649      & 0.898 & 0.693  & 0.684          & 0.652 & 0.713   \\
SyncoBERT     & 0.722 & 0.677      & 0.913 & 0.724  & 0.723          & 0.678 & 0.740   \\
UniXcoder     & 0.732 & 0.669      & 0.907 & 0.720  & 0.711          & 0.656 & 0.733   \\
CodeRetriever & 0.771 & 0.719      & 0.924 & 0.758  & \textbf{0.765} & 0.708 & 0.774   \\
CoCoSoDa      & 0.802 & 0.757      & 0.921 & 0.736  & 0.754          & 0.696 & 0.778   \\ \hline
CPLCS &
  \textbf{\begin{tabular}[c]{@{}l@{}}0.812\end{tabular}} &
  \textbf{\begin{tabular}[c]{@{}l@{}}0.757\end{tabular}} &
  \textbf{\begin{tabular}[c]{@{}l@{}}0.928\end{tabular}} &
  \textbf{\begin{tabular}[c]{@{}l@{}}0.764\end{tabular}} &
  \begin{tabular}[c]{@{}l@{}}0.760\end{tabular} &
  \textbf{\begin{tabular}[c]{@{}l@{}}0.715\end{tabular}} &
  \textbf{0.789} \\ \hline
\end{tabular}
\end{table*}

\textbf{RQ1: How does our approach perform compared to other state-of-the-art code search methods?}
We evaluate the effectiveness of our approach by comparing it to other state-of-the-art (SOTA) models introduced in Section 4.1. We reproduced CodeBERT, GraphCodeBERT, UniXcoder, Zecoler and CoCoSoDa in our own experimental setup. Regarding SyncoBERT and CodeRetriever, since they do not have open-source code, we directly use the results reported in the respective papers. The experimental results are presented in Table \ref{tab:2}. We have included the results for other metrics in the Appendix, and the conclusions that apply to MRR also extend to these metrics. Based on the results shown in the table, we can see that GraphCodeBERT outperforms CodeBERT in the two early pre-trained models (the first row of Table \ref{tab:2}) because it takes the data flow information of the code into account.

The five contrastive trained models (the second row of Table \ref{tab:2}) perform better than the two early models, which shows the effectiveness of the contrastive learning. Among them, Zecoler, which is based on CodeBERT but incorporates prompt tuning, outperforms GraphCodeBERT. CodeRetriever, which benefits from a larger training corpus, achieves better results than SyncoBERT, UniXcoder, and Zecoler. The best performing model, CoCoSoDa, presents a contribution to soft data enhancement techniques. It should be noted that the results we obtained from reproducing the effects of CoCoSoDa and UniXcoder are not as high as those reported in the original papers. We attribute this difference to the memory limitations of our experimental equipment, because we can only train with a smaller batch size. In contrastive learning, a larger batch size usually leads to better results.

\begin{figure}[htbp]
  \centering
  \includegraphics[width=\linewidth]{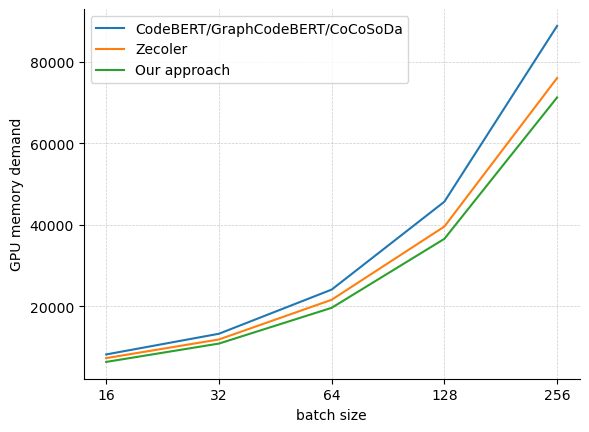}
  \caption{Comparison of GPU memory demands for different methods. Because the encoder structures of CodeBERT, GraphCodeBERT and CoCoSoDa are identical, their demand for GPU memory follows the same curve, which is depicted as a blue line in the figure.}
  \label{fig:3}
\end{figure}

Overall, our model performs the best except for the Java corpus, and it remains to be the most efficient one due to its lower resource consumption and faster training speed. As demonstrated in Figure \ref{fig:3}, by utilizing prompt learning during training, Zecoler and our CPLCS require only approximately 80\% GPU memory of the fine-tuning methods for a fixed batch size. Additionally, our method exhibits lower GPU memory requirements compared to Zecoler. In our experimental environment, CPLCS has only 19\% of the trainable parameters of the fine-tuning methods, and trains approximately 22\% faster than the fine-tuning methods.

\textbf{RQ2: How effective are the main components of CPLCS?}
In this subsection, we implement the ablation experiments to analyze the effectiveness of CPLCS from the viewpoints of different components: prompt tuning and cross-modal interaction mechanism. Regarding the experiment settings, we run four groups of experiments: i) DE+FT, which consists of the pure dual-encoder structure model trained by contrastive fine-tuning; ii) DE+FT+interaction, which includes a model with a cross-modal interaction mechanism added to a dual-encoder trained by contrastive fine-tuning; iii) DE+PT, using contrastive prompt tuning instead of fine-tuning to train dual-encoder model; iv) DE+PT+interaction, which utilizes the cross-modal interaction mechanism, in addition to the dual-encoder model trained by contrastive prompt tuning.

From Table \ref{tab:3}, we can observe that the performance of the model declines after removing any component. This means that each component plays an important role in our CPLCS. Prompt tuning can improve the quality of representation through contrastive learning, while the cross-modal interaction mechanism preserves fine-grained similarity modeling. As a result, both components are enhanced in the dual-encoder setup. The mechanisms act on orthogonal principles, leading to better results in setting iv) than those in setting ii) or iii). We have included the results for other PLs in the Appendix, and the conclusions that apply to Python also extend to other PLs.

\begin{table}[htbp]
\centering
\caption{Ablation study (Python)}
\label{tab:3}
\begin{tabular}{lllll}
\hline
Model                     & MRR   & R@1      & R@5      & R@10 \\ \hline
DE+FT                     & 0.736 & 0.633    & 0.865    & 0.920     \\
DE+FT+interaction         & 0.748 & 0.644    & 0.881    & 0.925     \\
DE+PT                     & 0.741 & 0.639    & 0.875    & 0.923     \\
DE+PT+interaction (CPLCS) & \textbf{0.764} & \textbf{0.669} & \textbf{0.894} & \textbf{0.935}     \\ \hline
\end{tabular}
\end{table}

\textbf{RQ3: Compared with other methods, how much improvement does our method make in semantic representation?}
There are two commonly used metrics for measuring the quality of a semantic representation: alignment and uniformity \cite{19}. They can be computed using the following equations:

\begin{equation}\label{equation_18}
\mathcal{L}_{align}=\underset{(X,Y) \sim p_{pos}}{\mathbb{E}}\left \| f(X)-f(Y)\right \|^2_2
\end{equation}

\begin{equation}\label{equation_19}
\mathcal{L}_{uniform}=log \underset{(X,Y) \stackrel{i.i.d}{\sim} p_{data}} {\mathbb{E}} e^{-2\left \| f(X)-f(Y)\right \|^2_2}
\end{equation}
where $(X,Y) \sim p_{pos}$ means $X$ and $Y$ are positive samples of each other, while $(X,Y) \stackrel{i.i.d}{\sim} p_{data}$ means that $X$ and $Y$ are independently and identically distributed; $f(X)$ and $f(Y)$ are the learned snippet-level representations and $\left \| f(X)-f(Y)\right \|^2_2$ represents the 2-norm of the distance between them. In the code search task, $\mathcal{L}_{align}$ is used to measure the alignment degree between a code-text positive sample pair. A lower value of $\mathcal{L}_{align}$ indicates a closer relationship between related code snippet and the query, resulting in a more accurate alignment. On the other hand, $\mathcal{L}_{uniform}$ indicates the distribution consistency of all code snippets or queries. A lower $\mathcal{L}_{uniform}$ suggests a more uniform distribution of the learned representations. The smaller the values of $\mathcal{L}_{align}$ and $\mathcal{L}_{uniform}$ in the corresponding higher-dimensional semantic space corresponding to a model, the better the PL and NL representations are learned, and the better overall performance generally.

\begin{figure}[htbp]
  \centering
  \includegraphics[width=\linewidth]{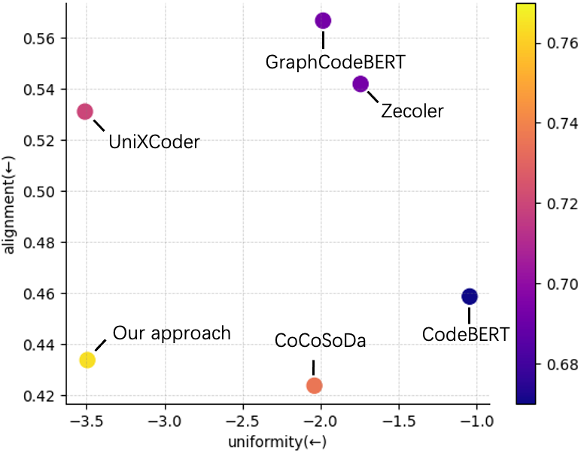}
  \caption{Representational quality map (Python)}
  \label{fig:4}
\end{figure}

In Figure \ref{fig:4}, we present the representation quality of six models in the Python corpus. The results reveal that CodeBERT demonstrates good alignment properties, but lacks uniformity. In contrast, GraphCodeBERT and Zecoler exhibit good consistency, but suboptimal alignment. Neither of these earlier methods achieve remarkable MRR scores. On the other hand, UniXCoder demonstrates significantly improved uniformity and performance. CoCoSoDa exhibits superior alignment, although slightly less consistency. Finally, our method is approximately 75\% more uniform than CoCoSoDa, while slightly less aligned, yielding the best results.

\textbf{RQ4: What is the Impact of Different Hyperparameters?}
In this section, we study the impact of different hyperparameters on the python language: length of prompt tokens, temperature hyperparameter $\tau$, interaction weight $\lambda$, and the maximum length of the input sequence (including PL and NL).

\begin{figure} [htbp]
    \centering
    \subfloat[\label{fig:5a}]{
	\includegraphics[scale=0.27]{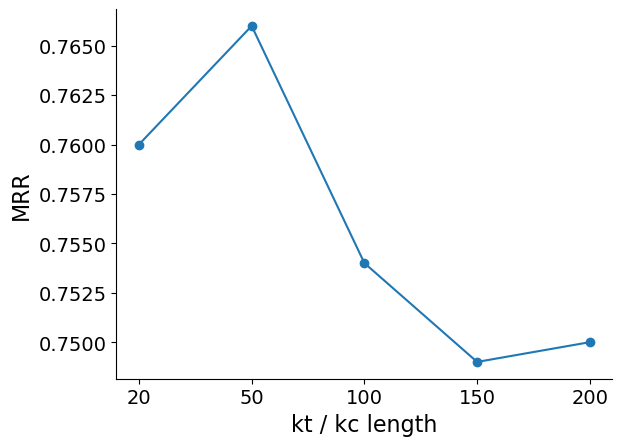}}
    \subfloat[\label{fig:5b}]{
	\includegraphics[scale=0.27]{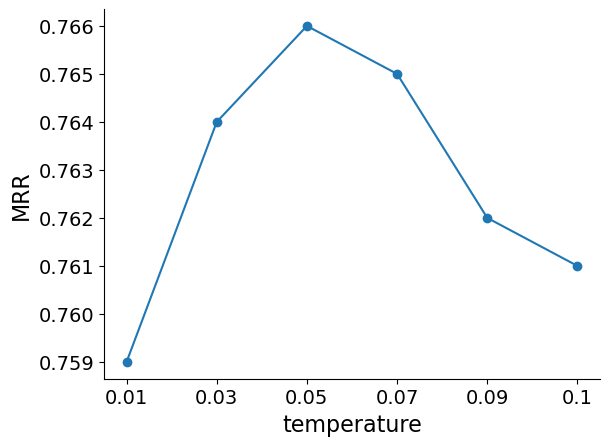}}
    \\
    \subfloat[\label{fig:5c}]{
	\includegraphics[scale=0.27]{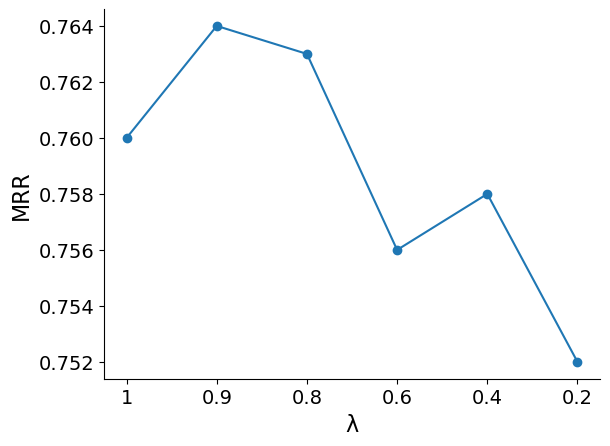}}
    \subfloat[\label{fig:5d}]{
	\includegraphics[scale=0.27]{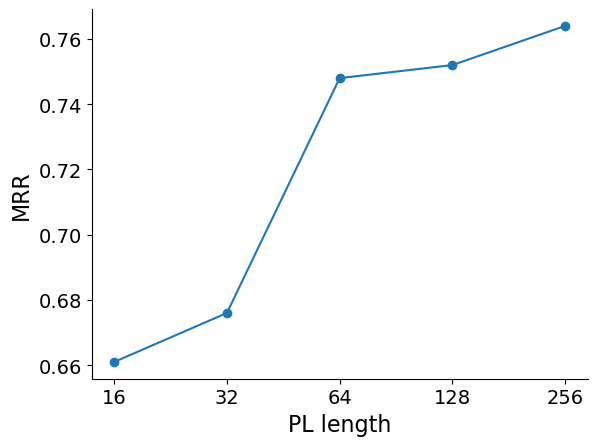}}
    \\
    \subfloat[\label{fig:5e}]{
	\includegraphics[scale=0.28]{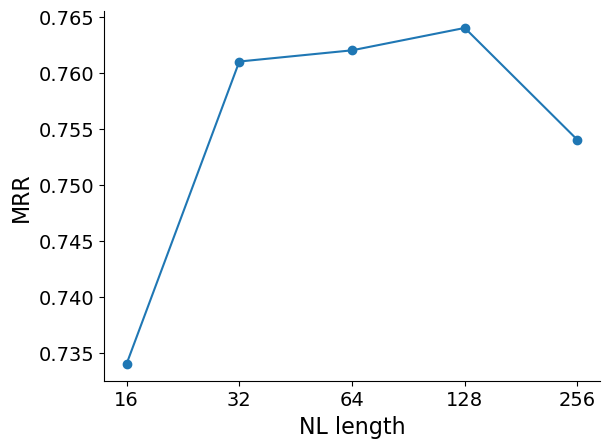}}
    \caption{Impact of different hyperparameters.}
    \label{fig:5} 
\end{figure}

Due to limited computing resources, we made a compromise in tuning hyperparameters, specifically allowing the prompt length for PL and NL to be the same, denoted as $kt = kc$. Referring to the experience of \cite{5,10,21}, we explore the effects of different hyperparameters within typical ranges. The experimental results are presented in Figure \ref{fig:5}. Our findings indicate that prompt length has a significant impact on the model's performance. The optimal results are obtained when the prompt length is approximately 50 tokens. Prompt length that is either too short or too long will adversely affect the performance of models (Figure.\ref{fig:5}\subref{fig:5a}). Temperature, as another key hyperparameter, exhibits a narrow stable range between 0.03 and 0.07, and excessively high or low temperature can lead to considerable performance degradation (Figure \ref{fig:5}\subref{fig:5b}). Setting the value of $\lambda$ to 0.9 results in peak performance of the model, while the performance maintains relatively stable within the range of 0.8 to 1.0 (Figure \ref{fig:5}\subref{fig:5c}). This finding suggests that the information contained in $u_{col}$ is more crucial than the information contained in $u_{row}$. In other words, the information contained in NL carries greater significance than that contained in PL. Based on the experiments conducted with different maximum input lengths for PL and NL, we observe that MRR is more sensitive to the changes in PL length compared to NL. The optimal PL length is found to be 256 (Figure \ref{fig:5}\subref{fig:5d}), while NL performed best at 128 (Figure \ref{fig:5}\subref{fig:5e}).

\section{Conclusion}
We have presented both experimental and analytical evidences to validate the efficacy of our CPLCS, an approach to address the issues of inadequate semantic representation and semantic gap between PL and NL. Experiments on a large real-world dataset across six languages have demonstrated the efficacy of our approach in enhancing representation quality and mapping ability between PL and NL. The experimental results indicate that our approach significantly outperforms other baseline methods.

\bibliographystyle{ACM-Reference-Format}
\bibliography{custom}

\appendix

\section{Performance Evaluation}
\textbf{RQ1: How does our approach perform compared to other state-of-the-art code search methods?}
We evaluate the effectiveness of our approach CPLCS by comparing it to several baselines on the CodeSearchNet dataset with six programming languages. We present the results for R@1, R@5 and R@10 here. The results are shown in Table \ref{tab:A1}, \ref{tab:A2} and \ref{tab:A3}, respectively. The results show that our approach performs best among all approaches. And because SyncoBERT and CodeRetriever models are not open source, there is no R@k metric data in the tables. 

\begin{table}[H]
\caption{Performance of different approaches on CodeSearchNet (R@1)}
\label{tab:A1}
\resizebox{\columnwidth}{!}{
\begin{tabular}{llllllll}
\hline
Model         & Ruby  & JavaScript & Go    & Python & Java  & PHP   & Overall \\ \hline
CodeBERT      & 0.582 & 0.514      & 0.835 & 0.574  & 0.578 & 0.525 & 0.601   \\
GraphCodeBERT & 0.605 & 0.543      & 0.853 & 0.594  & 0.593 & 0.545 & 0.622   \\ \hline
Zecoler       & 0.601 & 0.549      & 0.855 & 0.591  & 0.589 & 0.549 & 0.622   \\
SyncoBERT     & -     & -          & -     & -      & -     & -     & -       \\
UniXcoder     & 0.635 & 0.571      & 0.862 & 0.617  & 0.613 & 0.556 & 0.642   \\
CodeRetriever & -     & -          & -     & -      & -     & -     & -       \\
CoCoSoDa      & 0.702 & 0.655      & 0.871 & 0.633  & 0.655 & 0.599 & 0.685   \\ \hline
Our approach  & 0.712 & 0.661      & 0.883 & 0.669  & 0.664 & 0.617 & 0.701   \\ \hline
\end{tabular}
}
\end{table}

\begin{table}[H]
\caption{Performance of different approaches on CodeSearchNet (R@5)}
\label{tab:A2}
\resizebox{\columnwidth}{!}{
\begin{tabular}{llllllll}
\hline
Model         & Ruby  & JavaScript & Go    & Python & Java  & PHP   & Overall \\ \hline
CodeBERT      & 0.802 & 0.752      & 0.946 & 0.790  & 0.797 & 0.750 & 0.806   \\
GraphCodeBERT & 0.827 & 0.770      & 0.953 & 0.812  & 0.820 & 0.783 & 0.828   \\ \hline
Zecoler       & 0.834 & 0.775      & 0.957 & 0.817  & 0.815 & 0.786 & 0.830   \\
SyncoBERT     & -     & -          & -     & -      & -     & -     & -       \\
UniXcoder     & 0.873 & 0.789      & 0.960 & 0.840  & 0.828 & 0.781 & 0.845   \\
CodeRetriever & -     & -          & -     & -      & -     & -     & -       \\
CoCoSoDa      & 0.931 & 0.882      & 0.969 & 0.865  & 0.880 & 0.810 & 0.884   \\ \hline
Our approach  & 0.942 & 0.878      & 0.981 & 0.894  & 0.885 & 0.834 & 0.902   \\ \hline
\end{tabular}
}
\end{table}

\begin{table}[H]
\caption{Performance of different approaches on CodeSearchNet (R@10)}
\label{tab:A3}
\resizebox{\columnwidth}{!}{
\begin{tabular}{llllllll}
\hline
Model         & Ruby  & JavaScript & Go    & Python & Java  & PHP   & Overall \\ \hline
CodeBERT      & 0.855 & 0.812      & 0.958 & 0.853  & 0.850 & 0.812 & 0.857   \\
GraphCodeBERT & 0.872 & 0.834      & 0.970 & 0.862  & 0.867 & 0.829 & 0.872   \\ \hline
Zecoler       & 0.875 & 0.836      & 0.972 & 0.861  & 0.866 & 0.831 & 0.874   \\
SyncoBERT     & -     & -          & -     & -      & -     & -     & -       \\
UniXcoder     & 0.923 & 0.851      & 0.977 & 0.885  & 0.876 & 0.841 & 0.892   \\
CodeRetriever & -     & -          & -     & -      & -     & -     & -       \\
CoCoSoDa      & 0.964 & 0.932      & 0.984 & 0.920  & 0.923 & 0.872 & 0.932   \\ \hline
Our approach  & 0.975 & 0.930      & 0.992 & 0.935  & 0.932 & 0.880 & 0.940   \\ \hline
\end{tabular}
}
\end{table}

\textbf{RQ2: How effective are the main components of CPLCS?}
In the main text, we showcase ablation study in the Python corpus. Here we list ablation studies for the rest of the programming languages. The results are shown in Table \ref{tab:A4}-\ref{tab:A8}.

\begin{table}[H]
\centering
\caption{Ablation study (Ruby)}
\label{tab:A4}
\resizebox{\columnwidth}{!}{
\begin{tabular}{lllll}
\hline
Model                       & MRR   & R@1 & R@5 & R@10 \\ \hline
DE+FT                       & 0.802 & 0.702    & 0.931    & 0.964     \\
DE+FT+interaction           & 0.809 & 0.710    & 0.932    & 0.971     \\
DE+PT                       & 0.806 & 0.706    & 0.935    & 0.969     \\
DE+PT+interaction (CPTCS)   & 0.812 & 0.712    & 0.942    & 0.975     \\ \hline
\end{tabular}}
\end{table}

\begin{table}[H]
\caption{Ablation study (JavaScript)}
\label{tab:A5}
\resizebox{\columnwidth}{!}{
\begin{tabular}{lllll}
\hline
Model                       & MRR   & R@1 & R@5 & R@10 \\ \hline
DE+FT                       & 0.757 & 0.655    & 0.882    & 0.932     \\
DE+FT+interaction           & 0.758 & 0.659    & 0.881    & 0.932     \\
DE+PT                       & 0.757 & 0.656    & 0.882    & 0.930     \\
DE+PT+interaction (CPTCS)   & 0.757 & 0.661    & 0.878    & 0.930     \\ \hline
\end{tabular}}
\end{table}

\begin{table}[H]
\caption{Ablation study (Go)}
\label{tab:A6}
\resizebox{\columnwidth}{!}{
\begin{tabular}{lllll}
\hline
Model                       & MRR   & R@1 & R@5 & R@10 \\ \hline
DE+FT                       & 0.921 & 0.871    & 0.969    & 0.984     \\
DE+FT+interaction           & 0.925 & 0.879    & 0.976    & 0.990     \\
DE+PT                       & 0.923 & 0.874    & 0.974    & 0.987     \\
DE+PT+interaction (CPTCS)   & 0.928 & 0.883    & 0.981    & 0.992     \\ \hline
\end{tabular}}
\end{table}

\begin{table}[H]
\caption{Ablation study (Java)}
\label{tab:A7}
\resizebox{\columnwidth}{!}{
\begin{tabular}{lllll}
\hline
Model                       & MRR   & R@1 & R@5 & R@10 \\ \hline
DE+FT                       & 0.754 & 0.655    & 0.880    & 0.923     \\
DE+FT+interaction           & 0.757 & 0.660    & 0.876    & 0.931     \\
DE+PT                       & 0.755 & 0.657    & 0.882    & 0.927     \\
DE+PT+interaction (CPTCS)   & 0.760 & 0.664    & 0.885    & 0.932     \\ \hline
\end{tabular}}
\end{table}

\begin{table}[H]
\caption{Ablation study (PHP)}
\label{tab:A8}
\resizebox{\columnwidth}{!}{
\begin{tabular}{lllll}
\hline
Model                       & MRR   & R@1 & R@5 & R@10 \\ \hline
DE+FT                       & 0.696 & 0.599    & 0.810    & 0.872     \\
DE+FT+interaction           & 0.704 & 0.611    & 0.826    & 0.877     \\
DE+PT                       & 0.701 & 0.608    & 0.819    & 0.877     \\
DE+PT+interaction (CPTCS)   & 0.715 & 0.617    & 0.834    & 0.880     \\ \hline
\end{tabular}}
\end{table}

\begin{figure} [h]
    \centering
    \subfloat[\label{fig:8a}][Ruby]{
	\includegraphics[width=0.5\columnwidth]{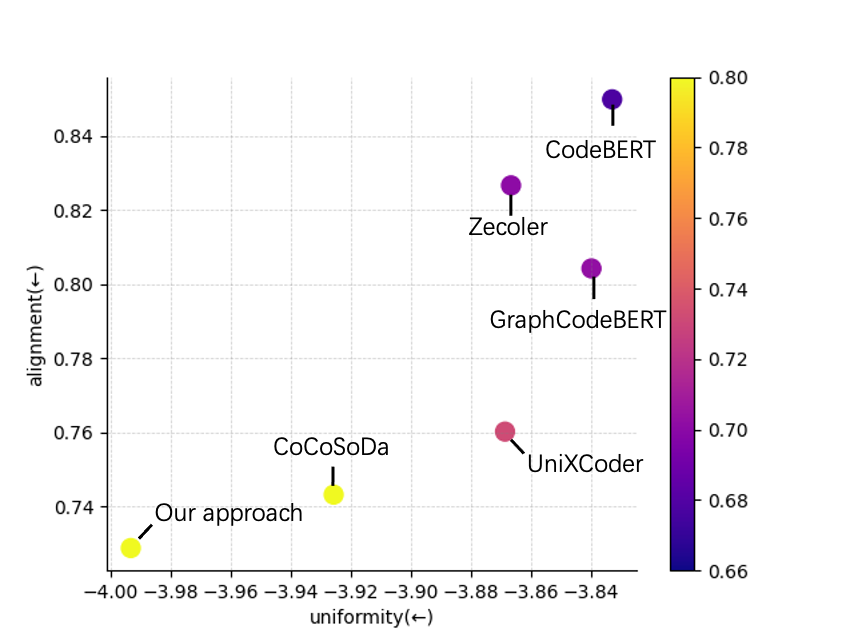}}
    \subfloat[\label{fig:8b}][JavaScript]{
	\includegraphics[width=0.5\columnwidth]{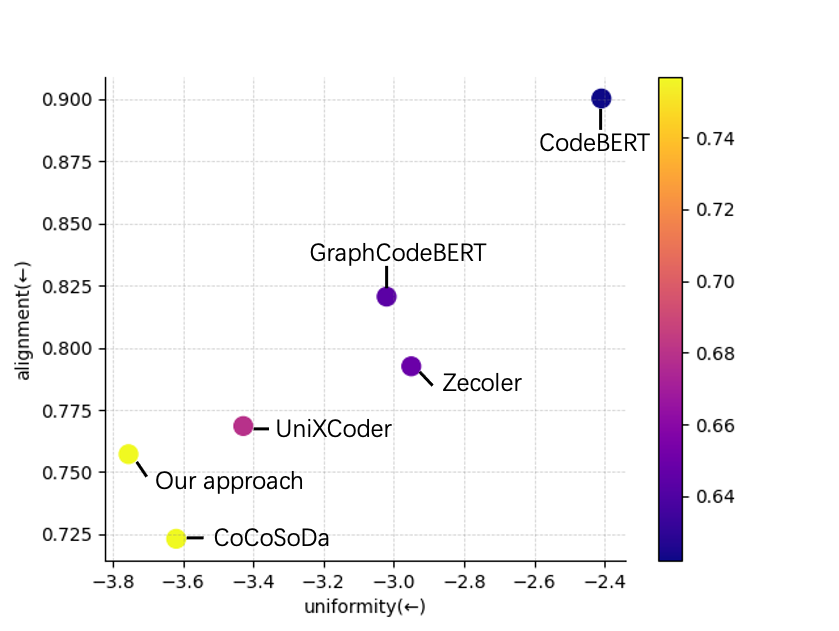}}
    \\
    \subfloat[\label{fig:8c}][Go]{
	\includegraphics[width=0.5\columnwidth]{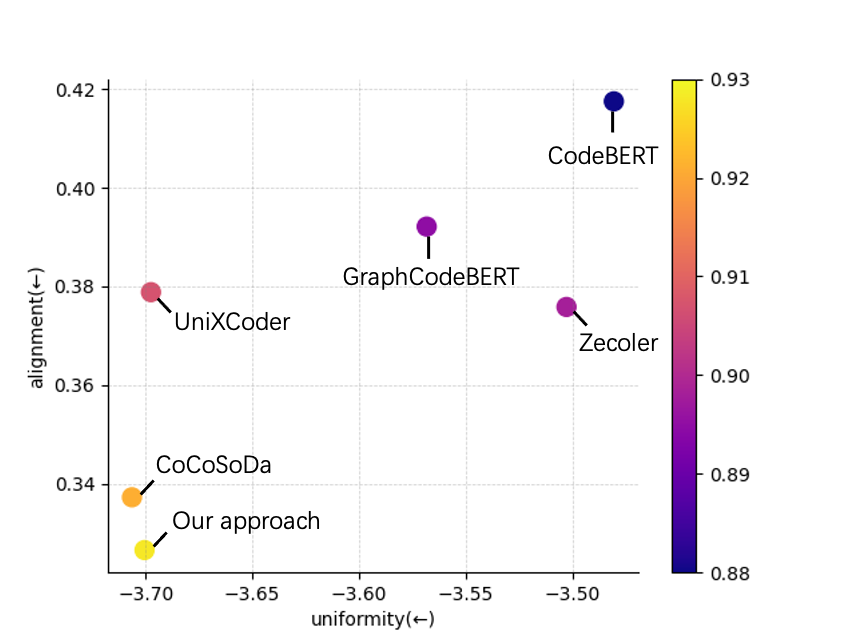}}
    \subfloat[\label{fig:8d}][Java]{
	\includegraphics[width=0.5\columnwidth]{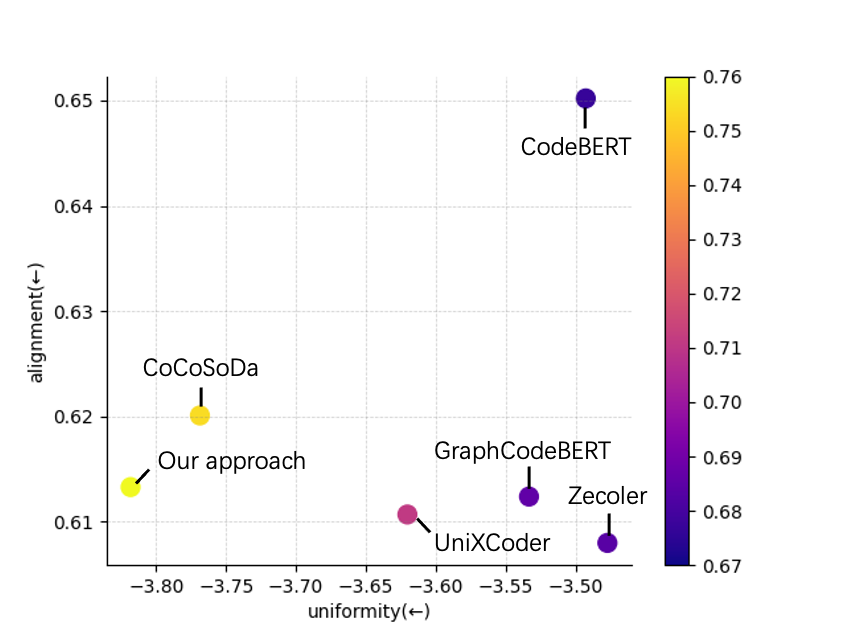}}
    \\
    \subfloat[\label{fig:8e}][PHP]{
	\includegraphics[width=0.5\columnwidth]{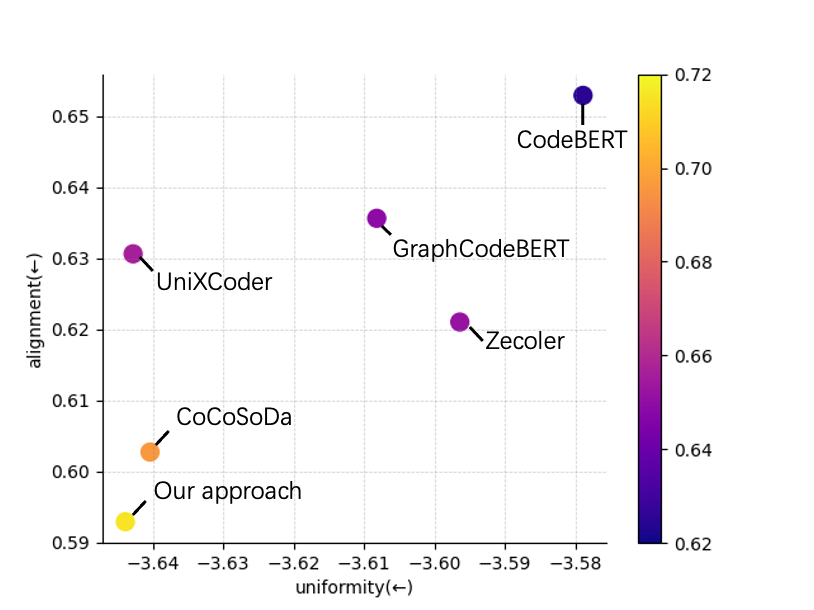}}
    \caption{ Representational quality map.}
    \label{fig:8} 
\end{figure}

\textbf{RQ3: Compared with other methods, how much improvement does our method make in semantic representation?}
In the main text, we showcase representation quality of several models in the Python corpus. Here we list representation quality maps for the rest of the programming languages. From the figure \ref{fig:8}, we can see that our model representation quality is the best among other languages.

\section{Case Study}
\begin{figure} [H]
    \centering
    \subfloat[\label{fig:6a}][The top 1 result of our approach]{
	\includegraphics[width=0.95\columnwidth]{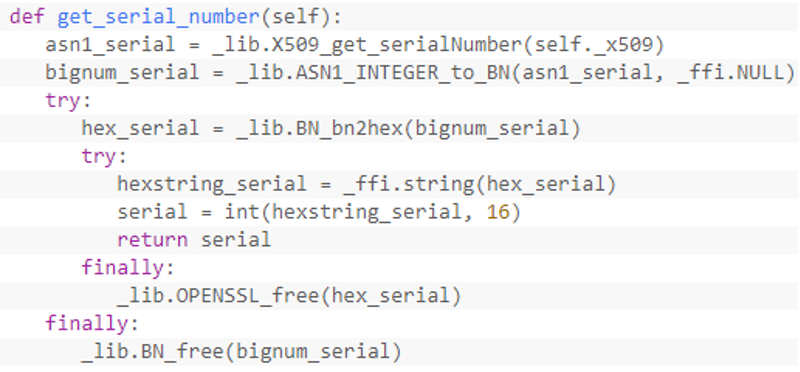}}
    \\
    \subfloat[\label{fig:6b}][The top 1 result of CoCoSoDa]{
	\includegraphics[width=0.95\columnwidth]{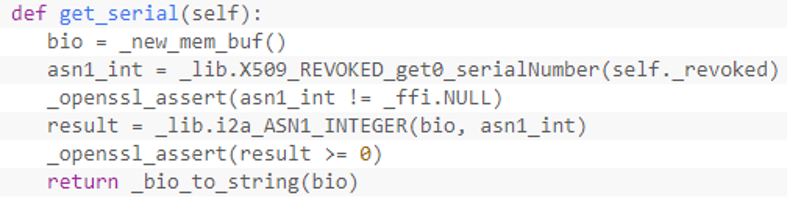}}
    \caption{The top 1 search result of our approach and CoCoSoDa for the query "Return the serial number of this certificate" on Python language}
    \label{fig:6} 
\end{figure}

Figure \ref{fig:6} shows the top 1 search result of our approach and CoCoSoDa for the query " Return the serial number of this certificate". Figure \ref{fig:6}(a) illustrates that our approach is capable of retrieving the ground truth code snippet. This is because our method features prompt tuning and a post-interaction mechanism, which allows for a more accurate understanding of the query and precise matching. On the other hand, Figure \ref{fig:6}(b) shows that CoCoSoDa’s retrieved code addresses the structure of revocation of the X.509 certificate, instead of the certificate itself.

\begin{figure} [H]
    \centering
    \subfloat[\label{fig:7a}][The top 1 result of our approach]{
	\includegraphics[width=0.95\columnwidth]{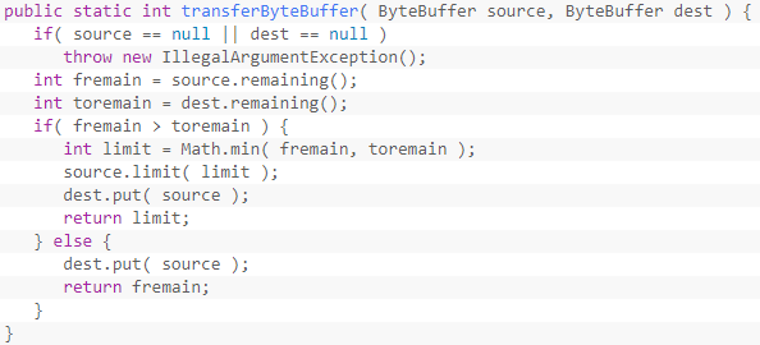}}
    \\
    \subfloat[\label{fig:7b}][The top 1 result of CoCoSoDa]{
	\includegraphics[width=0.95\columnwidth]{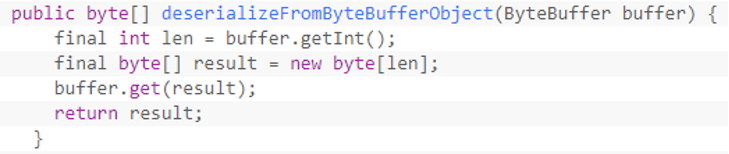}}
    \caption{The top 1 search result of our approach and CoCoSoDa for the query "Transfer from one ByteBuffer to another ByteBuffer" on Java language}
    \label{fig:7} 
\end{figure}

Figure \ref{fig:7} shows another search result for the query "Transfer from one ByteBuffer to another ByteBuffer". Obviously, from Figure \ref{fig:7}(a), we can see that the code snippets returned by our method transfer data from one ByteBuffer to another ByteBuffer through the put method, and return the number of bytes transferred, which aligns with the query intent. In contrast, Figure \ref{fig:7}(b) shows that CoCoSoDa is misled by a similar deserialization operation.

\end{document}